\documentclass[twocolumn,showpacs,preprintnumbers,amsmath,amssymb]{revtex4}
\usepackage{amsmath}
\usepackage{bm}
\usepackage{epsfig}
\usepackage[colorlinks=true,linkcolor=blue]{hyperref}

\newcommand{\be}{\begin{equation}}
\newcommand{\ee}{\end{equation}}

\newcommand{\bea}{\begin{eqnarray}}
\newcommand{\eea}{\end{eqnarray}}

\newcommand{\al}{\alpha}
\newcommand{\bt}{\beta}

\newcommand{\dl}{\delta}

\newcommand{\et}{\eta}

\newcommand{\lm}{\lambda}

\newcommand{\rh}{\rho}

\newcommand{\ph}{\phi}

\newcommand{\om}{\omega}

\newcommand{\rarrow}{\rightarrow}
\newcommand{\Rarrow}{\Rightarrow}

\newcommand{\nn}{\nonumber}

\begin{document}

\title{Dynamo effects in magnetized ideal-plasma cosmologies}

\author{Kostas Kleidis$^{1,2}$, Apostolos Kuiroukidis$^{1,3}$, Demetrios Papadopoulos$^1$ and Loukas Vlahos$^1$}

\affiliation{$^1$Department of Physics, Aristotle University of
Thessaloniki, 54124 Thessaloniki, Greece}

\affiliation{$^2$Department of Civil Engineering, Technological
Education Institute of Serres, 62124 Serres, Greece}

\affiliation{$^3$Department of Informatics, Technological
Education Institute of Serres, 62124 Serres, Greece}

\date{\today}

\begin{abstract}
The excitation of cosmological perturbations in an anisotropic
cosmological model and in the presence of a homogeneous magnetic
field has been studied, using the ideal magnetohydrodynamic (MHD)
equations. In this case, the system of partial differential
equations which governs the evolution of the magnetized
cosmological perturbations can be solved analytically. Our results
verify that fast-magnetosonic modes propagating normal to the
magnetic field, are excited. But, what's most important, is that,
at late times, the {\em magnetic-induction contrast} $(\dl B / B)$
grows, resulting in the enhancement of the ambient magnetic field.
This process can be particularly favored by condensations, formed
within the plasma fluid due to gravitational instabilities.
\end{abstract}


\maketitle

\section{Introduction}

Magnetic fields are known to have a widespread presence in our
Universe, being a common property of the intergalactic medium in
galaxy clusters~\cite{1}, while, reports on Faraday rotation imply
significant magnetic fields in condensations at high redshifts
~\cite{2}. Large-scale magnetic fields and their potential
implications for the formation and the evolution of the observed
structures, have been the subject of continuous theoretical
investigation in the past~\cite{3} -~\cite{6}. It became clear
that if magnetism has a cosmological origin, it could have
affected the evolution of the Universe~\cite{7}. Today, there are
several scenarios for the generation of primordial magnetic
fields~\cite{8}. The majority of the recent studies use a
Friedmann - Robertson - Walker (FRW) model to represent the
evolving Universe and super-impose a large-scale magnetic
field~\cite{9}. In other words, the magnetic field is assumed to
be too weak to destroy the FRW isotropy and any potential
anisotropy induced by it, is treated as a perturbation~\cite{10}.

However, mathematically speaking, the spatial isotropy of the FRW
Universe is not compatible with the presence of large-scale
magnetic fields. In fact, an anisotropic cosmological model can
and should be imposed for the treatment of magnetic fields whose
{\em coherent length} is comparable to the horizon
length~\cite{11}. Therefore, although current observations give a
strong motivation for the adoption of a FRW model, the effects one
may lose by neglecting the large-scale anisotropy induced by the
background magnetic field, should be investigated. Not to mention
that the anisotropy of the unperturbed model facilitates a closer
study of the coupling between magnetism and geometry.

In a recent paper~\cite{12}, which hereafter is referred to as
Paper I, the evolution of a magnetized, resistive plasma in an
anisotropic space-time has been studied numerically. The
corresponding results suggested that fast-magnetosonic waves grow
steeply with time and saturated at high values, due to the
resistivity. Nevertheless, numerics indicated also that growing
modes are present even in the limit of zero resistivity (the {\em
ideal plasma} case). Therefore, in the present article we consider
the same model, but, this time in the limit of the ideal MHD
approximation; namely, the assumption that the magnetic field is
frozen into an effectively infinitely conductive cosmic medium (a
fluid of zero resistivity). As we find out, in this case, the
evolution of the cosmological perturbations can be treated
analytically.

Following the procedure described in Paper I, we begin with a
resistive plasma, driving the dynamics of an anisotropic
space-time, in the presence of a time-dependent magnetic field.
This dynamical system is subsequently perturbed by small-scale
fluctuations and we study their interaction with the curved
background, searching for imprints on the temporal evolution of
the perturbations' amplitude. In particular:

In Section II, we present the system of the field equations
describing the model under consideration and the corresponding
zeroth-order solution. Accordingly, in Section III, we extract the
first-order perturbed equations and, confining ourselves in the
limit of zero-resistivity, we derive an analytic solution for the
magnetized cosmological perturbations. Our results verify that,
fast-magnetosonic modes are excited within the ideal plasma. At
late times, the magnetic induction contrast $(\dl B / B)$ grows,
resulting in the enhancement of the ambient magnetic field (dynamo
effect). This effect is particularly favored by condensations that
can be formed within the anisotropic fluid, due to gravitational
instabilities.

\section{A magnetized anisotropic Cosmology}

We consider an axially-symmetric Bianchi-Type I cosmological
model, driven by an anisotropic and resistive perfect fluid, in
the presence of a time-dependent magnetic field, ${\vec B} = B(t)
{\hat x}$. In the system of units where $\hbar = 1 = c$, the
corresponding line-element is written in the form \be ds^2 = -
dt^2 + R^2 (t) dx^2 + S^2 (t) [dy^2 + dz^2] \ee where, the
dimensionless scale factors $R(t)$ and $S(t)$ are functions of the
time coordinate.

The evolution of a curved space-time in the presence of matter and
an e/m field, is determined by the gravitational field equations
\be {\cal R}_{\mu \nu} - {1 \over 2} g_{\mu \nu} {\cal R} = 8 \pi
G {\cal T}_{\mu \nu} \ee together with the energy-momentum
conservation law \be {\cal T}_{ \; \; ; \nu}^{\mu\nu} = 0 \ee and
Maxwell's equations \bea F_{\; \; ;\nu}^{\mu \nu} & = & 4 \pi J^{\mu}, \\
F_{\mu \nu ; \lm} & + & F_{\nu \lm ; \mu} + F_{\lm \mu ; \nu} = 0
\eea In Eqs (2) - (5), Greek indices refer to the four-dimensional
space-time (in accordance, Latin indices refer to the
three-dimensional spatial section) and the semicolon denotes
covariant derivative. Furthermore, ${\cal R}_{\mu\nu}$ and ${\cal
R}$ are the Ricci tensor and the scalar curvature with respect to
the background metric $g_{\mu \nu}$, while $G$ is Newton's
gravitational constant. Eventually, $F^{\mu \nu}$ is the
antisymmetric tensor of the e/m field and $J^{\mu}$ is the
corresponding current-density.

The energy-momentum tensor involved, consists of two parts;
namely, \be {\cal T}^{\mu \nu} = {\cal T}_{fluid}^{\mu \nu} +
{\cal T}_{em}^{\mu \nu}\ee The first part, is due to an
anisotropic perfect-fluid source of the form \be {\cal T}^{\mu
\nu}_{fluid}= \rh u^0 u^0 + p_i u^i u^i + p_i g^{ii} \ee where,
$\rh(t)$ is the energy-density, $p_i(t)$ are the components of the
anisotropic pressure and the axial symmetry of the metric (1)
implies that $p_2 (t) = p_3 (t)$. In Eq (7), $u^{\mu} =
dx^{\mu}/ds$ is the fluid's four-velocity, satisfying the
conditions $u_{\mu} u^{\mu} = -1$ and $h^{\mu \nu} u_{\mu} = 0$,
with $h^{\mu \nu} = g^{\mu \nu} + u^{\mu} u^{\nu}$ being the
projection tensor.

The second part of ${\cal T}^{\mu \nu}$, is due to the ambient e/m
field \be {\cal T}^{\mu \nu }_{em}= {1 \over 4 \pi} ( F^{\mu \al}
F^{\nu \bt} g_{\al \bt} - \frac{1}{4} g^{\mu \nu} F_{\al \bt}
F^{\al \bt}) \ee In the absence of electric fields (something that
is strongly suggested by Paper I), the only non-zero components of
the Faraday tensor in the curved space-time (1) are \be F^{23}=
{B^x \over S^2} = -F^{32} \ee As regards the current-density
$J^{\mu}$, it can be determined by the invariant form of Ohm's law
\be J^{\mu} = n_e e u^{\mu} + {1 \over \et} F^{\mu \nu} u_{\nu}
\ee where, $n_e$ is the locally measured charge-density and $\et$
is the electric resistivity, in units of {\it time}. It is
reasonable to assume that, locally, the fluid has zero net-charge.
In this case, Eq (10) reduces to $J^{\mu} = {1 \over \et} F^{\mu
\nu} u_{\nu}$ and the identity $J_{; \mu}^{\mu} = 0$ holds, as a
consequence of the Maxwell equations. We have to point out that,
in the zero-resistivity limit Eq (4) reduces, through Eq (10), to
\be F^{\mu \nu} u_{\nu} = 0 \ee [clf Eq (B2) of Paper I]. In other
words, for $\et = 0$, we are left only with the {\em convective
field}~\cite{13}.

Following the procedure described in detail in Paper I, we look
for an axially-symmetric Bianchi-Type I cosmological solution to
the Einstein-Maxwell equations (2) - (5), representing the
zeroth-order solution of our problem. In this case, Eqs (2) reduce
to \bea &&2 \left(\frac{\dot{R}\dot{S}}{R S}\right) +
\left(\frac{\dot{S}}{S}\right)^{2} = 8 \pi G
\rh(t) + G B^2(t)\nn \\
&&-2\frac{\ddot{S}}{S} - \left(\frac{\dot{S}}{S}\right)^{2}
= 8 \pi G p_1(t) - G B^2(t)\\
&&-\left(\frac{\ddot{R}}{R} + \frac{\ddot{S}}{S}\right)
-\left(\frac{\dot{R}\dot{S}}{R S}\right) = 8 \pi G p_2(t) + G
B^2(t) \nn \eea where, the dot denotes time-derivative. For ${\vec
E} = {\vec 0}$ and ${\vec B} (t)$ // ${\hat x}$, Eq (4) vanishes
identically and Eq (5) yields \be \partial_{t}[S^2 B(t)] = 0 \; \;
\Rarrow \; B(t) = {B_0 \over S^2} \ee where, $B_0$ is the initial
value of the magnetic induction. Eq (13) has a clear physical
interpretation: The magnetic flux through a comoving surface
normal to the direction of the magnetic field, {\em is conserved}.
Furthermore, the continuity equation (3) results in \bea
&&\partial_t [\rh(t) + {B^2 \over 8 \pi}] + {\dot{R} \over R} \: [
p_1(t) - {B^2 \over 8 \pi}] + 2 {\dot{S}
\over S} \: [ p_2(t) + {B^2 \over 8 \pi} ] \nn \\
&&+ ({\dot{R} \over R} + 2 {\dot{S} \over S}) \: [\rh (t) + {B^2
\over 8 \pi}] = 0 \eea and the particles' number conservation law,
reads \be {\dot{\rh}} + (\frac{\dot{R}}{R} + 2 \frac{\dot{S}}{S})
\rho = 0 \; \; \Rarrow \; \rh(t) = {\rh_0 \over RS^2} \ee where,
$\rh_0$ is the initial energy-density. In Paper I, we showed that
the system of Eqs (12) - (15) admits the exact solution \bea R(t)
= ({t \over t_0}), \; \; \; \; S(t) = ({t \over t_0})^{1 \over 2}
\nn \\ \rh (t) = \rh_0 ({t_0 \over
t})^2 \; \; \; \; B (t) = B_0 ({t_0 \over t}) \\
p_1 (t) = p_{10} ({t_0 \over t})^2 , \; \; \; p_2 (t) = - p_{20}
({t_0 \over t})^2 \nn \eea where, the index "$0$" stands for the
corresponding values at $t = t_0$, which marks the beginning of
the interaction between magnetized plasma and curved space-time.
Solution (16), describes the evolution of an anisotropic
cosmological model, in which, the large-scale anisotropy along the
$\hat{x}$-axis, is due to the presence of an ambient magnetic
field. The combination of Eqs (12) and (16) indicates that,
initially, the equation of state for the matter-energy content
reads \be p_0 = {1 \over 5} (\rho_0 + 6 {B_0^2 \over 8 \pi}) \ee
and, therefore, as regards the perfect fluid itself, we obtain
$p_0 = {1 \over 5} \rh_0$. This {\em soft equation of state}
enlists the curved space-time (1) among the {\em semi-realistic}
cosmological models of Bianchi Type I. These models are crude,
first-order approximations to the actual Universe, when we use
currently available theories and observations~\cite{14}.

\section{Cosmological perturbations in an ideal plasma fluid}

For every dynamical system, much can be learnt by investigating
the possible modes of small-amplitude oscillations or waves. A
plasma is physically much more complicated than an ideal gas,
especially when there is an externally applied magnetic field. As
a result, a variety of small-scale perturbations may appear.

Following the formalism of Paper I, we introduce first-order
perturbations to the Einstein-Maxwell equations, by decomposing
the physical variables of the fluid as \bea \rho (t,z) & = & \rho
(t) + \dl \rho(t,z) \\
p_x (t,z) & = & p_1 (t) \nn \\
p_y (t,z) & = & p_2 (t) - \dl p (t,z) \\
p_z (t,z) & = & p_2 (t) + \dl p (t,z) \nn \eea and we insert the
perturbed values (18) and (19) into Eqs (12) - (15), neglecting
all terms higher or equal than the second order. The pressure
perturbation $\dl p (t, z)$ introduces a longitudinal acoustic
mode, propagating along the ${\hat z}$-direction \be \dl p (t, z)
= C_s^2 \: \dl \rh (t, z) \ee where, $C_s = {1 \over \sqrt {5}}$
is the speed of sound. The four-velocity of the plasma fluid is
perturbed around its comoving value, $u^{\mu} = (1,0,0,0)$, as \be
u^{\mu} (t,z) = (1+\dl u^{0}(t,z),0,0,\dl u^{z}(t,z)) \ee Then,
the condition $u_{\mu} u^{\mu} = -1$, to the first leading order,
implies \be \dl u^{0}(t,z)=0 \ee and, therefore, $u^3(t, z) = \dl
u^z (t, z)$. Accordingly, $\rho (t,z)u^3(t,z) = \rho (t) \dl
u^z(t,z) + O_2$.

As regards the perturbations of the e/m field, we consider that
they correspond to a transverse e/m wave, propagating along the
${\hat z}$-axis $({\vec k} // {\hat z})$; namely, \bea
\vec{E}(t,z) & = &\dl E^{y}(t,z)\hat{y} \\
\vec{B}(t,z)&=&B(t)\hat{x}+\dl B^{x}(t,z)\hat{x} \eea Therefore,
now, the non-zero components of the Faraday tensor in the curved
space-time (1) are modified as follows \bea && F^{02}= {1 \over S}
\: \dl E^y (t, z) = - F^{20} \nn \\
&& F^{23} = {1 \over S^2} \: [B(t) + \dl B^x(t, z)] = - F^{32}
\eea In the search for dynamo effects within the linear regime, we
should stress that magnetic fields (as well as their
inhomogeneities) are not created by first-order terms in the
metric perturbations: Their production involves electric currents
generated by the rotational component of the velocity of the
plasma constituents, along the lines of the so-called {\em
vorticity effect}~\cite{15},~\cite{16} and this component arises
only at the second-order approximation~\cite{17} -~\cite{22}.

Within the limits of linear analysis, excitation of cosmological
perturbations in a homogeneous and anisotropic cosmological model
is basically a {\em kinematic effect}, in the sense that the
self-gravitation of the fluctuations is unimportant (e.g.
see~\cite{23}, pp. 501 - 506 and references therein). In this
case, perturbations' growth may arise mostly due to their motion
in the anisotropic background.

Therefore, as far as the enhancement of MHD perturbations in the
anisotropic space-time (1) is concerned, we may neglect the first
order corrections of the metric, admitting the so-called {\em
Cowling approximation}~\cite{24}. In other words, in what follows
we treat the MHD perturbations as very low-frequency
(small-amplitude) waves propagating in an anisotropically
expanding medium, without interacting with the curved space-time
unless the linear regime breaks down. Accordingly, the evolution
of the perturbed quantities is governed only by the
energy-momentum tensor conservation, together with Maxwell's
equations.

The linearly-independent, first-order perturbed Einstein-Maxwell
equations in the curved background (1), first extracted in Paper
I, read \bea &&\partial_t (\dl \rho ) + \rho(t)
\partial_z (\dl u^z) + \dl \rho \left(\frac{\dot{R}}{R} +
2\frac{\dot{S}}{S}\right) = 0 \\ &&-\partial_t (\dl E^{y}) + {1
\over S} \partial_z (\dl B^x) - \dl E^{y} \left (
\frac{\dot{R}}{R} + \frac{\dot{S}}{S} \right) = \nn
\\ &&= 4 \pi {1 \over \et} [\dl E^{y} + S B (t) \dl u^{z}] \eea
\be \partial_{t}(S^2 \dl B^{x}) - S \partial_{z} (\dl E^{y}) = 0
\ee \bea &&\partial_{t} [\rho(t)\dl u^{z} - {1 \over 4 \pi S} B(t)
\dl E^{y}] + {1 \over S^2} \partial_{z}
[\dl p + {1 \over 4 \pi} B(t) \dl B^{x}]+ \nn \\
&&+\left(\frac{\dot{R}}{R} + 2\frac{\dot{S}}{S}\right) [\rho(t)
\dl u^{z} - {1 \over 4 \pi S} B(t) \dl E^{y}]=0 \eea Confining
ourselves in an ideal plasma, Eq (27) results in \be \dl E^y = - S
B(t) \dl u^z \ee verifying that, for $\et = 0$, one is left only
with the perturbations of the {\em convective field}.
Differentiating Eq (30) with respect to $z$ \be \partial_z (\dl
E^y) = - S B(t) \partial_z (\dl u^z) \ee and taking into account
the conservation of the magnetic flux, Eq (28) is written in the
form \be \partial_t (S^2 \dl B^x) = - B_0 \partial_z (\dl u^z) \ee
On the other hand, with the aid of the particles' number
conservation law, Eq (26) reads \be \partial_z (\dl u^z) = - {1
\over \rh_0} \: \partial_t [RS^2 (\dl \rh)] \ee and therefore, Eq
(32) yields \be \partial_t (S^2 \dl B^x) = {B_0 \over \rh_0} \:
\partial_t [ RS^2 (\dl \rh)] \ee We integrate Eq (34) with respect
to $t$, to obtain \be \dl B^x = {B_0 \over \rh_0} \: R(t) \: \dl
\rh + {1 \over S^2} f(z) \ee Without loss of generality, we may
impose that $f (z) = 0$. In fact, in an expanding Universe, the
term $f(z)/S^2$ eventually becomes negligible due to the
cosmological redshift. Therefore, as regards the temporal
evolution of $\dl B^x$, to set $f(z) = 0$ simply corresponds to a
translation in the origin of time, to the point where $f(z)/S^2$
vanishes. Accordingly, \be \dl B^x = {B_0 \over \rh_0} \: R(t) \:
\dl \rh \ee With the aid of Eqs (13) and (15), Eq (36) results in
\be {\dl B \over B (t)} = {\dl \rh \over \rh (t)} \ee We observe
that, in an anisotropic cosmological model which is driven by
ideal plasma, the {\em magnetic induction contrast} is equal to
the {\em energy-density contrast}. In other words, the magnetic
field perturbation amplifies in tune with the energy-density
perturbation and therefore, any Jeans instability (condensations
that can be formed inside the plasma fluid due to an unstable
growth in $\dl \rh$) results in the increase of $\dl B^x$ and,
hence, acts in favor of {\em dynamo effects}. In fact, Eq (37) was
first obtained in~\cite{11}, regarding a Bianchi Type I model
filled with ideal plasma, within the context of the so-called {\em
covariant formalism} (e.g. see Eq (98) of ~\cite{11}). Coincidence
of these results, although in~\cite{11} perturbations of the
metric were taken into account, justifies the assumption that the
evolution of the cosmological perturbations in an anisotropic
background is a purely kinematic effect~\cite{23}.

Now, the system of the first-order perturbation equations (26) -
(28) is written in the form \bea && \dl B^x = {B_0 \over \rh_0} \:
R(t) \: \dl \rh \nn \\ &&\partial_z (\dl u^z) = - {1 \over \rh_0}
\: \partial_t [RS^2 (\dl \rh)] \\ &&
\partial_z (\dl E^y) = {1 \over S} \: ({B_0 \over \rh_0}) \:
\partial_t [RS^2 (\dl \rh)] \nn \eea where, the last equation has
resulted from the combination of Eqs (31) and (33), with the aid
of the magnetic flux conservation. From Eqs (38) it becomes
evident that the behavior of the velocity and the electric field
perturbations is modulated mostly due to the inherent anisotropy
of the expanding medium. We also note that the evolution of every
perturbation quantity on the lhs, involves the evolution of $\dl
\rh$. The first-order perturbation of the energy-density is
accordingly determined by Eq (29), which, upon consideration of
Eqs (30) and (36), reduces to \be \rh_0
\partial_t [ (1 + u_A^2 {R \over S^2}) \dl u^z] + R (C_s^2 + u_A^2
{R \over S^2})
\partial_z (\dl \rh) = 0 \ee where, $u_A^2 = B_0^2 /4 \pi \rh_0$
is the (dimensionless) Alfv\'{e}n velocity. However, as regards
the cosmological model under study, we have $R(t) = S^2(t)$ and,
therefore, Eq (39) takes on its final form \be \rh_0 (1 + u_A^2)
\partial_t (\dl u^z) + R (C_s^2 + u_A^2) \partial_z (\dl \rh) = 0
\ee We differentiate Eq (33) with respect to $t$, to obtain \be
\partial_t \partial_z (\dl u^z) = - {1 \over \rh_0} \:
\partial_t^2 [RS^2 (\dl \rh)] \ee Accordingly, we differentiate Eq
(40) with respect to $z$, to obtain \be \rh_0 (1 + u_A^2)
\partial_z \partial_t (\dl u^z) + R(t) \: (C_s^2 + u_A^2)
\partial_z^2 (\dl \rh) = 0 \ee The combination of Eqs (41) and
(42) results in \be - (1 +u_A^2) \partial_t^2 \Psi (t, z) + {1
\over S^2} \: (C_s^2 + u_A^2) \partial_z^2 \Psi (t, z) = 0 \ee
where, we have set \be \Psi (t, z) = R S^2 (\dl \rh) \ee To solve
Eq (43), we use the method of separation of variables, considering
\be \Psi (t, z) = T(t) Z(z) \ee Now, the equation which governs
the evolution of $\Psi (t, z)$, is written in the form \be {1 +
u_A^2 \over C_s^2 + u_A^2} \: ({t \over t_0}) \: {1 \over T(t)} \:
{\partial^2 \over \partial t^2} T(t) = {1 \over Z(z)} \:
{\partial^2 \over \partial z^2} Z(z) \ee implying that both parts
are equal to an arbitrary constant $(\lm)$; namely, \be
{\partial^2 \over \partial z^2} Z(z) = \lm Z(z) \ee and \be
{\partial^2 \over \partial t^2} T (t) = \lm \: ({t_0 \over t}) \:
{C_s^2 + u_A^2 \over 1 + u_A^2} \: T(t) \ee In Paper I we have
showed that the only perturbation modes admitted in an ideal
plasma are the magnetosonic waves \be (1 + u_A^2) \om_0^2 = (C_s^2
+u_A^2) k^2 \ee (clf Eq (48) of~\cite{12}) where, $k$ is the
comoving wave-number and $\om_0$ is the corresponding
angular-frequency of the wave. In order to incorporate these modes
in our analysis, we need to impose \be \lm = - k^2 \ee in which
case, Eqs (47) and (48) result in \be Z(z) \sim e^{i k z} \ee and
\be {\partial^2 \over \partial t^2} T (t) + \om_0^2 \: ({t_0 \over
t}) \: T(t) = 0 \ee Eq (52) admits formal Bessel-type
solutions~\cite{25} \be T(x) = \sqrt {x} \: [c_1 J_1 (2 \om_0 t_0
\sqrt {x}) + c_2 Y_1 (2 \om_0 t_0 \sqrt {x}) ] \ee where, $c_1$
and $c_2$ are arbitrary integration constants and we have set \be
1 \leq x = {t \over t_0} < \infty \ee Therefore, the
energy-density perturbation reads \be \dl \rh = {1 \over x^{3/2}}
\: [ c_1 J_1 (2 \om_0 t_0 \sqrt {x}) + c_2 Y_1 (2 \om_0 t_0 \sqrt
{x}) ] e^{i k z} \ee To find the functional form of the other
perturbation quantities, we use Eqs (38) and take into account the
following recurrence relations of the Bessel functions~\cite{25}
\bea &&J_{\nu}^{\prime} (\ph) = J_{\nu - 1} (\ph) - {\nu \over
\ph} J_{\nu} (\ph) \nn \\
&&Y_{\nu}^{\prime} (\ph) = Y_{\nu - 1} (\ph) - {\nu \over \ph}
Y_{\nu} (\ph) \eea where, a prime denotes differentiation with
respect to the argument $\ph$. Accordingly, we obtain \bea && \dl
u^z = {\om_0 \over k \rh_0} \: [ c_1 J_0 (2 \om_0 t_0 \sqrt {x}) +
c_2 Y_0 (2 \om_0 t_0 \sqrt {x}) ] e^{i (k z + {\pi \over 2})} \\
&& \dl E^y = {\om_0 B_0 \over k \rh_0 \sqrt {x}} \: [ c_1 J_0 (2
\om_0 t_0 \sqrt {x}) + c_2 Y_0 (2 \om_0 t_0 \sqrt {x}) ] e^{i (k z
- {\pi \over 2})} \\ && \dl B^x = {B_0 \over \rh_0 \sqrt{x}} \: [
c_1 J_1 (2 \om_0 t_0 \sqrt {x}) + c_2 Y_1 (2 \om_0 t_0 \sqrt {x})
] e^{i k z} \eea With the particular choice $c_2 = -i c_1$, the
time-dependent amplitude of the cosmological perturbations can be
written in the form of Hankel functions; namely, \bea && \dl \rh =
{1 \over x^{3/2}} \: c_1 \: H_1^{(2)} (2 \om_0 t_0 \sqrt {x}) \:
e^{i k z}
\\ && \dl u^z = {\om_0 \over k \rh_0} \: c_1 \: H_0^{(2)} (2 \om_0
t_0 \sqrt {x}) \: e^{i (k z + {\pi \over 2})} \\ && \dl E^y =
{\om_0 B_0 \over k \rh_0} \: {1 \over x^{1/2}} \: c_1 \: H_0^{(2)}
(2 \om_0 t_0 \sqrt {x}) \: e^{i (k z - {\pi \over 2})} \\ && \dl
B^x = {B_0 \over \rh_0} \: {1 \over x^{1/2}} \: c_1 \: H_1^{(2)}
(2 \om_0 t_0 \sqrt {x}) \: e^{i k z} \eea Eqs (60) - (63) possess
a very interesting asymptotic behavior at both large and small
arguments (e.g. see~\cite{26}).

Although it is quite clear what do we mean by the large-argument
behavior; it corresponds to what the perturbations' may look like
at late times $(t \rarrow \infty)$, the corresponding
small-argument limit is rather ambiguous, since $t \geq t_0$. At
this point we should emphasize that the magnetosonic waves are, in
fact, {\em low frequency modes}. Therefore, we expect that $\om_0$
should be much smaller than the characteristic frequency for the
matter and energy to shift about, along the direction of
propagation, at $t = t_0$; namely, the Hubble parameter across the
$z$-axis. Accordingly, \be \om_0 \ll H_S (t_0) = {1 \over 2 t_0}
\; \; \Rarrow \; 2 \om_0 t_0 \ll 1 \ee Therefore, the
small-argument behavior simply corresponds to small values of $t$
after $t_0$.

\subsection{Asymptotic behavior for small values of the argument}

In this case, which corresponds to an early-time approximation ($x
\simeq 1$ and $2 \om_0 t_0 \sqrt {x} \ll 1$), the cosmological
perturbations read \bea && \dl \rh = {c_1 \over
\pi \om_0 t_0} \: {1 \over x^2} \: e^{i (k z + {\pi \over 2})} \\
&& \dl u^z = {2 \om_0 \over \pi k \rh_0} \: c_1 \:
\ln {1 \over \om_0 t_0 \sqrt {x}} \: e^{i (k z + \pi)} \\
&& \dl E^y = {2 \om_0 B_0 \over \pi k \rh_0} \: c_1 \: {1 \over
\sqrt {x}} \: \ln {1 \over \om_0 t_0 \sqrt {x}} \: e^{i k z}
\\ && \dl B^x = {B_0 \over \rh_0} \: {c_1 \over \pi \om_0 t_0}
\: {1 \over x} \: e^{i (k z + {\pi \over 2})} \eea We observe
that, both the energy-density and the magnetic field perturbation
decrease adiabatically, due to the cosmological expansion [clf Eq
(16)]. The amplitude of the magnetic induction contrast is written
in the form \be \vert {\dl B \over B} \vert_{init} = {c_1 \over
\pi \om_0 t_0} \: {1 \over \rh_0} = \vert {\dl \rh \over \rh}
\vert_{init} \ee i.e. it is equal to the constant value of the
energy-density contrast [a not unexpected result, clf Eq (37)].

\subsection{Asymptotic behavior for large values of the argument}

In this case, which corresponds to a late-time approximation ($x
\rarrow \infty$ and $2 \om_0 t_0 \sqrt {x} \gg 1$), the
cosmological perturbations read \bea && \dl \rh = {1 \over
x^{7/4}} \: {c_1 \over \sqrt {\pi \om_0
t_0}} \: e^{i (k z - 2 \om_0 t_0 \sqrt {x} + {3 \pi \over 4})} \\
&& \dl u^z = {1 \over x^{1/4}} \: {\om_0 \over k \rh_0} \: {c_1
\over \sqrt {\pi \om_0
t_0}} \: e^{i (k z - 2 \om_0 t_0 \sqrt {x} + {3 \pi \over 4})} \\
&& \dl E^y = {1 \over x^{3/4}} \: {\om_0 B_0 \over k \rh_0} \:
{c_1 \over \sqrt {\pi \om_0 t_0}} \: e^{i (k z - 2 \om_0 t_0 \sqrt
{x} - {\pi \over 4})} \\ && \dl B^x = {1 \over x^{3/4}} \: {B_0
\over \rh_0} \: {c_1 \over \sqrt {\pi \om_0 t_0}} \: e^{i (k z - 2
\om_0 t_0 \sqrt {x} + {3 \pi \over 4})} \eea Now, the amplitude of
the magnetic induction contrast is written in the form \be \vert
{\dl B \over B} \vert_{final} = {c_1 / \rh_0 \over \sqrt {\pi
\om_0 t_0}} \; ({t \over t_0})^{1/4} \ee and the same relation
holds for the energy-density contrast, as well. It is worth noting
that, while initially $(\dl B / B)$ acquires a constant value, at
late times, it results in an increasing function of time. In fact,
combining Eqs (69) and (74), we obtain \be \vert {\dl B \over B}
\vert_{final} = \sqrt {\pi \om_0 t_0} \; ({t \over t_0})^{1/4} \:
\vert {\dl B \over B} \vert_{init} \ee Since both $(\dl B / B)$
and $(\dl \rh / \rh)$ are increasing functions of time, there is a
characteristic time, $t_c$, at which \be {\dl B \over B} \; , \;
{\dl \rh \over \rh} \simeq 1 \ee and the linear analysis breaks
down. However, the growth of $\dl B$ with respect to $B(t)$ is, in
fact, a very slow process. We may calculate explicitly the
temporal limits of the linear approach, in a realistic setting.

Small-angle anisotropy in the CMRB implies that, for adiabatic
perturbations along the recombination epoch $(t_R)$, one
has~\cite{27} \be ({\dl T \over T})_R = {1 \over 3} \: ({\dl \rh
\over \rh})_R \ee Accordingly, at $t = t_R$ we have \be \vert {\dl
B \over B} \vert_{final} = \vert {\dl \rh \over \rh} \vert_{final}
\lesssim 3 \: ( {\dl T \over T} )_R \ee where, the second step
follows because the MHD modes can constitute no more than 100\% of
the observed CMRB anisotropy. With the aid of Eq (78) we may
determine the arbitrary constant involved in Eq (74); namely \be
c_1 = 3 \: ( {\dl T \over T} )_R \: \rh_0 \: \sqrt { \pi \om_0
t_0} \: ( {t_0 \over t_R} )^{1/4} \ee and hence, in terms of
$t_R$, Eq (74) reads \be \vert {\dl B \over B} \vert_{final}
\lesssim 3 \: ( {\dl T \over T} )_R \: ( {t \over t_R} )^{1/4} \ee
We have to point out that, in Eq (80) the temporal limits are $t_0
\leq t \leq t_R$, since this mechanism does not apply after $t_R$.
In other words, no large-scale inhomogeneities of the magnetic
field can be formed after recombination, when electrons and
protons combine to form neutral hydrogen and radiation decouples
from matter (in connection, see also~\cite{21}).

Nevertheless, extrapolating this result beyond $t_R$, we may take
a good idea of how slow the enhancement of these magnetic
fluctuations may be. In this case, according to Eq (80), the
characteristic time at which $\dl B \simeq B$ and the linear
approach is no longer valid, reads \be t_c \gtrsim {1 \over 81
({\dl T \over T})_R^4 } \: t_R \ee Current observations (e.g.
see~\cite{28}) imply that \be ({\dl T \over T})_R \sim 10^{-5} \ee
and therefore \be t_c \gtrsim 10^{18} \: t_R \ee Notice that, even
if Eq (74) [or (80)] was valid also in the matter-dominated era,
the amplification of $\dl B$ should have been continued up to
$10^{18}$ times the recombination epoch ($10^{14}$ times the age
of the Universe) before it can become comparable to $B_0$.

Therefore, dynamo effects (enhancement of the ambient magnetic
field) do take place in an anisotropic background, with the linear
approach being valid at high accuracy. According to Eq (37), this
is most certainly true in the presence of large scale
condensations within the cosmic fluid.

\vspace{.3cm}

\section{Conclusions}

In the present article, we explore the possibility of enhancing a
primordial ({\em seed}) magnetic field, by taking advantage of the
large-scale anisotropy created by it, within a fluid of infinite
conductivity. To do so, we study the evolution of (low-frequency)
magnetosonic waves in an anisotropically expanding ideal plasma
and in the presence of a homogeneous magnetic field along the
$\hat{x}$-direction.

Following the procedure described in Paper I~\cite{12}, we begin
with a resistive fluid driving the dynamics of the curved
space-time (the so-called zeroth-order solution). This dynamical
system is subsequently perturbed by small-scale fluctuations and
we study their interaction with the anisotropic background,
searching for imprints on the temporal evolution of the
perturbations' amplitude.

Confining ourselves in the limit of zero resistivity (infinite
conductivity), we solve analytically the system of partial
differential equations which governs the evolution of the
magnetized cosmological perturbations. Our results verify that,
fast-magnetosonic modes are excited within the ideal plasma. But,
what's most important, is that, at late times, the magnetic
induction contrast $(\dl B / B)$ grows, following a power-law
temporal dependence, thus leading to the enhancement of the
ambient magnetic field (dynamo effect), with the linear
approximation being valid at high accuracy. This effect is
particularly favored by condensations that can be formed within
the anisotropic fluid, due to a gravitational instability, since,
as we recover, the magnetic induction contrast amplifies in tune
with the corresponding energy-density counterpart.

\vspace{.3cm}

The authors would like to thank Dr Christos G. Tsagas for helpful
discussions. We also thank the anonymous referee fir his critical
comments and his useful suggestions, which greatly improved this
article's final form. Finally, financial support from the Greek
Ministry of Education under the Pythagoras programm, is gratefully
acknowledged.

\end{document}